# Sorting Integers on the AP1000

Lex Weaver
&
Andrew Lynes

May 21, 1994

**Abstract**

Sorting is one of the classic problems of computer science. Whilst well understood on sequential machines, the diversity of architectures amongst parallel systems means that algorithms do not perform uniformly on all platforms. This document describes the implementation of an radix based algorithm for sorting positive integers on a Fujitsu AP1000 Supercomputer, which was constructed as an entry in the Joint Symposium on Parallel Processing (JSPP) 1994 Parallel Software Contest (PSC94). Brief consideration is also given to a full radix sort conducted in parallel across the machine.

# Contents

# 1 Introduction

Sorting is one of the classic problems of computer science. Algorithms for sorting on sequential machines are well understood and can be found in most standard texts. This is not true for parallel sorting techniques where much theoretical work has been done, but little of general applicability has been implemented. The reason for this is not only the relative scarcity of parallel machines, but also their great diversity. This variation in hardware gives rise to a situation in which the best algorithm for one platform may give vastly inferior performance on another. The investigation detailed here was dual purpose, it was entered with both the intention of gaining a fuller understanding of the AP1000 and how its architecture could be applied to the sorting problem, and also in the hope of generating a competitive entry in the JSPP[1]'94 Parallel Software Contest (PSC94). The requirements of this contest effectively determined the program specification which is discussed in §1.1. A discussion of the preliminary problem investigation follows, along with a description of the algorithm first implemented. Subsequent performance enhancing modifications are described in §4 with §5 considering an alternative algorithm prior to conclusions being drawn.

## 1.1 Program Specification

Since the program was to be an entry in PSC94, the specification was determined by the problem set for that contest. This was to sort up to 64 million positive 32bit integers using only 64 cells of an AP1000. The initial data distribution was unknown and after sorting, the integers were to be evenly distributed across the 64 cells. Cell 0 containing the lowest valued integers, cell 1 the next lowest, ..., and cell 63 the highest.

The contest organisers provided an object file containing functions which facilitated the generation of 5 distinct data sets, the checking of the sort result, and provided the official times. An analysis of the data sets quickly revealed that they were designed to provide a thorough test of any parallel integer sorting program. The following table details observations about each of the 5 keys, with *values* referring to the distribution of values within the possible range of positive integers, and *balance* being the distribution of elements across the cells.

| key | values | balance | comment |
| --- | --- | --- | --- |
| 1 | random | good | the best case |
| 2 | random | bad | to test balancing |
| 3 | uniform | good | to test naive implementations of qsort |
| 4 | some clustering | mildly unbalanced | a general test |
| 5 | clustered | good | to test bucketing/radix type methods |

---

[1] Joint Symposium on Parallel Processing

# 2 Preliminary Investigation

The AP1000 is a medium-grained MIMD distributed memory machine which utilises *wormhole routing* on a toroidal grid[1] connecting up to 1024 processors. The wormhole routing is hardware supported and permits two processes to communicate using the torodial grid network (T-net) without interrupting the intervening processors through which the message would otherwise have to pass. On a machine as small as the 64 cells being presently considered, varying inter-processor distances are negligible [2].

Parallel machines based on store and forward routing algorithms are usually based on the hypercube architecture to reduce the number of intervening processors, and hence minimise the cost of communication. Sorting algorithms for such machines tend to rely on nearest neighbour communication since this eliminates both intervening processors and the likelihood of network contention. Since it is well established that communication on the AP1000 is relatively cheap and contention rarely observed, it is not surprising that the work of Keating and Nelmes [2] revealed communication minimisation to be not an important factor when sorting on the AP1000. They implemented and compared several algorithms for a similar problem to that being considered, with what they term a *radix* sort proving the fastest. This algorithm involved a distribution phase in which all processors sent messages of varying sizes to all other processors. Such a communication pattern would be intolerable on a store and forward machine, yet was quicker than more traditional parallel algorithms such as the bitonic sort [5].

## 2.1 A General Purpose Parallel Sorting Algorithm

The work of Tridgell and Brent[9] in implementing a general purpose parallel sorting algorithm on the AP1000 was seen as highly relevant. Their algorithm essentially solved the required problem in the general case rather than specifically for integers. Several phases make up the algorithm they present;

### 2.1.1 Pre-Balancing

For the purpose of load balancing and to maintain the preconditions of the other phases this ensures that all processors have the same number of elements. It may be necessary to utilise *infinity padding*[2] if the total number of elements is not a multiple of the number of processors.

The algorithm is based upon a hypercube communication pattern, with each node balancing with each of its hypercube neighbours in turn. This method has two disadvantages. Firstly, on a 64 cell machine it forces each processor to at least negotiate (send and receive) with six other cells in a specific order, and secondly, if data is moved during the balance, it may be moved up to three times before reaching its destination cell. The advantage of the algorithm is that the total number of elements need not be known in advance, although on the AP1000 this could be determined quite cheaply.

---

[2]This involves creating extra elements which are easily identifiable, such as inf, and hence easily removeable after the sort has been completed.

### 2.1.2 Serial Sorting

This phase utilises a fast sequential algorithm to sort the elements local to each processor.

### 2.1.3 Primary Merging

The aim of this phase is to almost sort the elements across the machine in a very short time. The hypercube communication pattern is used once again, but with a *merge-exchange operation* between the processors as opposed to the simple data transfer as done in pre-balancing.

**Merge-Exchange Operation** — This is an operation between two processors that results in one processor possessing the lower valued half of the two processors' combined elements and the other the upper half.

### 2.1.4 Cleanup

The cleanup phase is actually capable of performing the entire sort and balancing itself, but is optimised to work on the almost sorted data which is generally the outcome of the previous phases. The algorithm is a generalisation of Batcher's merge-exchange algorithm and is described in [5].

## 2.2 Execution Times

The best performance for integer sorting reported in [9] was slightly less than 4*$10^6$ elements per second. Without integer optimisations the results were considerably poorer at $10^6$ elements per second. For comparison, an estimate of the execution time of a perfectly efficient 64 processor parallel machine running qsort() on $64*10^6$ integers was made. The figure calculated[3] was $\frac{64*10^6 \log_2(64*10^6)(10^6 \ elements \ seq. \ sort \ time)}{64*(10^6*\log(10^6))} \approx 88.76$ seconds, or about 225,000 elements per second.

With these times in mind it was decided that any parallel program geared specifically towards sorting positive integers on a 64 cell AP1000 would have to sort at least 4*$10^6$ elements per second, ie: completely sorting $64*10^6$ elements in approximately 16 seconds.

---

[3]qsort() is an implementation of the quicksort algorithm which is $O(n \log n)$.

# 3 The Initial Algorithm

After consideration of both [2] and [9] it was decided to implement an algorithm similar to that of Tridgell and Brent, but with the primary merging phase replaced with a bucket based distribution phase, which for simple elements such as integers on which bit-wise comparisons are valid, should be faster. The phases of the implemented algorithm are detailed below.

1. Pre-Balance
2. Bucketing and Distribution
3. Local Sort
4. Cleanup
5. Post-Balance

## 3.1 Pre-Balance

The pre-balance algorithm utilised makes use of the fact that each cell has available to it not only the number of elements it possesses but also the total number of elements. From these two figures each cell can calculate its own shortfall or excess of elements with respect to an even distribution, this is known as the cell's *delta*. Once each cell's delta is known it is possible to send excess elements directly to their final destination and thus avoid double or triple data movements.

The steps necessary to achieve this are;

1. calculate local delta
2. broadcast local delta and receive deltas from other cells
3. sort the delta's into ascending order, ie: from biggest shortfall to largest excess
4. distribute excesses

The broadcast step may appear to be expensive at first, consisting of 64 broadcasts, but using the low-latency hardware supported communication calls on the AP1000 these come relatively cheaply. Each call to xy_brd() costs in the order of $10^{-4}$ seconds so even 64 consecutive calls does not constitute a significant amount of time. If the algorithm were to be scaled up to more processors, it may become useful to replace this step with the hypercube global array summation as discussed in §4.

The distribution of excesses is the most interesting step of those above. The cells each fall into one of three categories, those with a zero local delta, those with a negative, and those with a positive. After the delta broadcasts and sorting, cells with a zero delta move on to the next phase immediately, and those with a negative delta receive elements from any sender until they have exactly the required number. Cells with a positive delta are required to do more work however. They must work through their sorted list of deltas, calculating how many cells will have their shortfalls satisfied by cells with greater excesses than themselves. This is quite simple to do since the larger excess are applied to the larger shortfalls first (and larger shortfalls consume the larger excesses first). Having determined which shortfalls will remain unsatisfied after all greater excesses have been consumed, the excess cell can determine what the final destinations[4] of its excess will be and proceed with the data transfer. An important point to note is that the sum of all deltas is zero, hence the sum of the excesses is the same magnitude as the sum of the shortfalls. Thus, there will always be somewhere for the excess to be sent, and always enough data received to make up the shortfall. The following pseudo-code details how a cell with positive delta determines the destinations of its excess.

```
if (local delta > 0) then
    while (top delta does not belong to this cell) do
        top delta = top delta + bottom delta;
        if (top delta < 0) then
            bottom delta = top delta;
        else
            increment bottom;
            if (top delta == 0) then
                decrement top;
            endif
        endif
    endwhile

    while (local delta > 0) do
        messagesize = min(local delta, bottom delta);
        send messagesize elements to bottom delta;
        local delta = local delta - messagesize;
        increment bottom;
    endwhile
endif
```

Two minor performance enhancements that this pre-balance algorithm benefits from are;

**Unordered Receiving** — previous experience with the AP1000 has shown that where a cell is to receive messages to be treated in the same manner from several senders, a noticeable[5] performance improvement can be gained by receiving the messages in no specific order. This most probably results from the situation where the transfer of a message that is to be received later under ordered receiving temporarily blocks the transfer of one that is being waited for. In the present case, where a cell continues to receive until its delta is zero, the necessity to communicate who is sending how much data is also removed.

---

[4]It is unlikely that there will be a one-to-one mapping between excesses and shortfalls, hence some excesses will be split up and others coalesced.

[5]In this context the meaning of noticeable should be taken as measurable rather than significant.

**Optimal Message size** — all elements transferred by the pre-balance are sent using a function called Big_Send() which breaks large messages up into smaller ones that occupy communication channels for a shorter length of time. Experimentation showed that the optimal standard size for a message was 64 000 elements, with even variations to 63 000 or 65 000 being noticeable.

## 3.2 Bucketing and Distribution

With regard to performance this is probably the most important phase of the algorithm. It seeks to quickly shift all of the elements to the cell which will be their final destination. This is attempted via a bucketing scheme which is very similar to the last pass of a radix sort (see §3.3). The following pseudo code describes the top level of the bucketing and distribution algorithm.

> **bucket(var** element[ ], **var** current_size)
>
> > **for** i = 0 **to** (current_size - 1) **do**
> > > increment bucket to which element[i] belongs;
> >
> > **endfor**
> > calculate cumulative totals for local buckets;
> > copy local bucket cumulative totals;
> > calculate cumulative totals for buckets globally;
> > partial sort(element[], current_size,
> > > local cumulative totals);
> >
> > distribute data(element[], current_size, global cumulative totals,
> > > local cumulative totals, local buckets);

The algorithm is implemented such that the number of buckets used is assumed to be a power of two. This enables simple and quick bitwise operations to determine which bucket an element belongs to. The entire range of positive integers is covered by the buckets used, and the logarithm to the base two of this number determines how many high order bits from the integer are needed to make up an integer's bucket index. Obviously the greater the number of buckets the smaller the range covered by any one bucket (or the better the bucketing resolution). This results in a generally better distribution of data amongst the cells, there is however a tradeoff in that the computation of cumulative totals, both local and global, will require both more time and space. The global cumulative totals were initially calculated using an xy_isum() operation for each bucket. Whilst these operations are individually very cheap, performing one for each bucket does take a considerable amount of time. The optimal number of buckets was determined by trial and error to be either 4096 or 8192, with memory requirements from the cleanup phase initially forcing the choice to be 4096.

The *partial sort* used above shuffles elements such that all those falling within the same bucket are grouped together, and that bucket groups are in ascending order. This is necessary for the efficient distribution of data, allowing large contiguous sections of the array holding the elements to be sent, rather than multiple small buckets.

The data distribution function uses the bucket totals provided to determine which sections of the partially sorted array should be sent to each processor, and proceeds with the sending. Since every other processor is doing the same, it then receives, into the array it just sent from, elements from other processors. A conscious decision was made to provide the distribution function with only the global cumulative totals across the buckets and not the cumulative totals across the processors for each bucket. It was felt that the communication time required to provide this extra information would be excessive. The cost of this decision was simply a small error in the total number of elements sent to some processors during distribution. It was estimated that this error should not exceed 128 elements in magnitude, and was in fact never observed to exceed eight. This slight imbalance was to be rectified by the post-balance phase. Another minor consequence of this method was a difficulty in terminating unordered receiving, which in the context of an all-to-all distribution phase with a minimum of 4096 messages traversing the network would most probably come into its own. The problem was eventually circumvented by accepting all elements until each of the 64 cells had been received from (NULL messages were sent if no elements were to be, and a cell would send data to itself) and the number of elements received was within a predetermined range about the required number of elements. This required a minor modification to the `Big_Send()`[6] function so that the last component message of a larger message would be greater in size than the predetermined range. Without this modification it would be possible for the last message of a multi-part send to be left unread, as the receiving processor would have already received from all processors and would have sufficient elements to be within the specified range.

The first, rather naive, implementation of the distribution function suffered a little from network contention with all processors sending to all others beginning with cell 0. Some staggering would have resulted from minor differences in the computation of earlier phases, but in general all processors were sending to the same processor at once. Changing this ordering so that each processor began by sending to itself and then moved up through the processor ids modulo 64 saw a measurable but by no means significant improvement. Adding a further offset of ten to this staggering, so that each processor effectively took a short break from transmission whilst sending to itself after ten other transmissions saw a significant drop (approximately two seconds) in performance on key 2, with marginal improvements on the others. No explanation other than an unusual distribution of data on key 2 causing a highly contentious sending pattern could be found for this behaviour, and the change was quickly abandoned.

## 3.3 Local Sort

There were 3 obvious candidates for the local serial sort. These were;

- `qsort()` from <stdlib.h> (an implementation of quicksort)
- `gnu_qsort()` as modified by Andrew Tridgell [9]
- a radix sort [5, 4]

---

[6]See §3.1

The first two had the advantages of being already implemented and available, whilst the third promised better performance for the large numbers of elements being considered. Implementation initially began using `qsort()` as the default, with a more informed decision to be made later. With some data distributions generated during development requiring in excess of five minutes for a local sort, it soon became apparent that `qsort()` was inadequate. Tridgell's modifications of `gnu_qsort()` consist of optimisations for integers that do not alter the time complexity of the algorithm, hence it was concluded that it did not warrant investigation. If `qsort()` couldn't cope, a slightly faster version of the same algorithm would still not be fast enough. It was thus decided to investigate a radix sort.

### 3.3.1 Radix Sort

The radix sort is closely related to the bucketing process described in §3.2. Essentially the data is "sorted" twice, firstly on the low order bits, then on the high order bits, with the low order "sort" being retained as the high order sort is performed. Further details can be found in [5], but the following two important features should be noted.

1. it is independent of data distribution, ie: its best, worst, and average cases are the same

2. it has *linear* time complexity

The linear time complexity yields great benefits over the $O(n \log_2(n))$ average case complexity of quicksort [4] on large data sets. When sorting 100,000 randomly distributed integers `qsort()` and the radix implementation were both comfortably under 5 seconds, but for $10^6$ integers `qsort()` required approximately 70 seconds whilst the radix was still under 10.

## 3.4 Cleanup

### 3.4.1 Motivation

The purpose of the cleanup phase was to ensure that the data were correctly sorted. The need for such a process resulted from the fact that the bucketing phase described previously provided a highly efficient although not necessarily accurate sorting algorithm. It was possible for any number of elements to have been sent to the incorrect cell. As a result, the output from all stages up to and including the local sort could not be guaranteed to be totally sorted, hence the need for a cleanup phase.

### 3.4.2 General Algorithm

Although the bucketing phase could not guarantee a complete sort, it was anticipated that the number of errors would be small compared to the size of the data set. In addition, it was expected that in the general case, a datum would be within one cell of its final destination. These two assumptions had a great impact on the cleanup algorithm that was used. Since it was assumed that the data were almost sorted, a communication pattern following Batcher's algorithm was deemed to be unnecessary. A simple linear algorithm was chosen instead. With this algorithm, a datum not located on the correct cell would be passed from one cell to another using a linear communication pattern until its final destination was reached. Obviously, if many elements fell into this category, or there were many elements that were more than one cell from their final destination, performance was disappointing.

The cleanup phase involved each cell comparing its minimum value with its left neighbour's maximum value. It follows therefore that each cell also compared its maximum value with its right neighbour's minimum value. The order in which the comparison took place was dependent on the cell ID. All cells whose cell ID was divisible by 2 participated in the comparison with their right neighbour before the comparison with their left neighbour. All other cells participated in the comparison with their left neighbour before moving on to their right neighbour. In the case of any overlap, the two cells would engage in a merge-exchange. At the end of each iteration, all cells participated in a global summation operation. All cells who had participated in a merge-exchange supplied a 0 to the summation, while all remaining cells supplied a 1. The algorithm terminated when the result of a global summation was equal to the total number of cells.

The merge-exchange algorithm involved two distinct stages. Initially, the two cells entered into a negotiation process where they determined the approximate number of elements to exchange in order to correctly perform the merge. The negotiation process began with each cell taking a sample of its local elements and exchanging it with the sample generated by its partner. Each sample element was taken to represent a fixed number of elements in the original list. Having obtained the sample elements from the neighbouring cell, both the local sample list and neighbouring sample list were traversed elementwise, until enough elements could be accounted for to ensure that the result of the merge would be correct. The required number of elements were then exchanged between the two cells. The algorithm also made use of dynamic load balancing. That is, at the end of merge process, both cells would have the same number of elements [7] stored locally.

There are two points to note about this negotiation method. Firstly, it is not accurate. In most cases, more data is exchanged than is necessary. However, in all cases, all necessary data is exchanged. The second point to note is that different quantities of data may be exchanged. That is, the amount of data being exchanged between the two cells may not necessarily be the same for both cells. This algorithm was used because the bucketing process described in §3.2 potentially leaves the cells with a data imbalance. As a result, it is difficult to exchange only the data necessary for use in the merge.

The second stage of the algorithm involves a merge between two lists. The merge algorithm is a variant of that generally used for such a purpose. Essentially memory is allocated to accommodate the final merged list. However, the final list does not contain all of the elements provided by the input lists. Instead, it contains only those elements which are to remain on the current cell. This relates to the fact that more data may be sent and received than necessary. Once the memory is allocated, the merge progresses in the classic manner with one notable exception. In this implementation, two different routines were used, depending on whether the lower values or higher values were to remain on the cell. The cell which is to keep the lower values will obviously call the first routine while the other cell will call the second routine. Once the merge has completed, the memory used to store the two source lists is freed and the merge list is used as the data list for the cell. At the end of the entire merge process, the two cells involved in the merge have the same number of elements in sorted order. In addition, the lower elements will be located on the cell with the lower cell ID, while the higher elements will be stored on the cell with the higher cell ID.

---

[7]In the case of an odd number of elements, the cell with the lower cell ID receives the extra datum

The merge algorithm described above performs well in terms of speed. However, the memory requirements of the algorithm are excessive. Having to supply memory for a target list as well as two source lists is the major limiting factor in the algorithm's use. Under the assumption that the bucketing algorithm would result in only a small number of elements to be placed on the incorrect cell however, it was expected that the memory limitation would not be observed.

The whole cleanup algorithm was based on the premise that the previous sorting methods would produce an almost sorted data set. As long as this premise was satisfied, the algorithm would be very fast. However, to violate this premise would produce very poor performance indeed.

## 3.5 Post-balance

### 3.5.1 Motivation

As a result of the bucketing process, there was a possibility that data would not be evenly distributed among the cells. The program specification required that each cell have the same number of elements at the end of the sorting process. In order to satisfy this condition, a post-balance phase was introduced.

### 3.5.2 General Algorithm

Unlike the pre-balance phase, the elements in the cells were assumed to be in sorted order. As a result, the post-balance could not destroy this ordering. This meant that any rebalancing should be implemented by generating a ripple effect. If a particular cell is in deficit, data would have to ripple from a cell or group of cells with a surplus of data items through all intermediate cells, to the cell with the deficit. In this way, the sorted order would be maintained.

Each cell had global variables representing the target size and current size of the data array. This information could be readily used to obtain a delta value, the number of elements the data array was in surplus or deficit. That is, $\Delta = current\_size - target\_size$. Once each cell had calculated delta, it was then broadcast to all other cells using `xy_brd()`. Only 64 broadcasts of 4 bytes were required in this case. As a result, it was unlikely that a more sophisticated communication pattern would achieve the same result in a lesser time. The cells used the received delta information to generate two values. The first represented the overall delta of all cells with a lesser cell ID, while the second represented the overall delta of all cells with a greater cell ID. This information could be used to generate the desired ripple effect.

If the cells with a lesser cell ID had an overall deficit, a cell would send enough data from the low end of its data array to satisfy the deficit. On the other hand, if the same cells had an overall surplus, the cell would receive the entire surplus into the lower end of its data array. If the cells with a higher cell ID had an overall deficit, a cell would once again send a enough data from the high end of its data array to satisfy the deficit. It follows therefore that in the case of a surplus in the cells with a higher cell ID, a cell would receive the entire deficit into the high end of its data array.

One possible flaw with this algorithm is that a cell may not be able to supply all the data necessary to cover a deficit. This situation could not occur in practice since the result of the bucketing and the cleanup processes described previously was a sorted and almost balanced collection of data.

# 4 The Quest for Speed.

## 4.1 Initial Performance

The performance of the algorithm described in the previous sections was excellent in most cases. Unfortunately, performance under worst case conditions was far from encouraging.

In Figure 1, the sorting time for each of the 5 keys is illustrated. It can be seen that for the first four distributions, sorting was completed in under 13 seconds. The performance for the fifth distribution however was substantially inferior, with the sorting process taking in excess of 40 seconds. Analysis of this distribution showed 25 % of the data to be clustered. In this situation, the bucketing stage was not able to obtain the resolution required in order to satisfactorily perform the global sort. As a result, the cleanup stage was required to perform a large amount of sorting, resulting in the poor performance. The poor performance was caused by the fact that the cleanup phase was designed to work efficiently on almost sorted data. Violating this pre-condition however, produced poor results.

The breakdown of the timing figures is also of interest. Figure 2 shows the total time consumed by each stage of the algorithm across all five data distributions. The five phases of the algorithm, numbered 1-5 in the figure, are pre-balance, bucketing, local sort, cleanup and post-balance. The figure clearly shows that the time taken for the pre-balance phase of the algorithm was insignificant compared to the other sections. The bucketing and local sort stages both required approximately the same amount of time. By far the most significant factor in the overall performance was the cleanup stage. It should be noted however that more than 30 seconds of the time used to perform the cleanup could be attributed to key 5. The post-balance phase of the algorithm consumed a notable although not significant amount of time.

## 4.2 CleanUp

In the original implementation, it was expected that the bucketing algorithm would produce an almost sorted data set. This was the basis of the entire cleanup algorithm. Unfortunately, this assumption was proven to be incorrect by the distribution generated by key 5. This was highlighted by the fact that the first four data distributions could each be sorted in under 13 seconds, while the fifth distribution required more than 40 seconds. It was apparent that a more comprehensive sorting algorithm was required in this case.

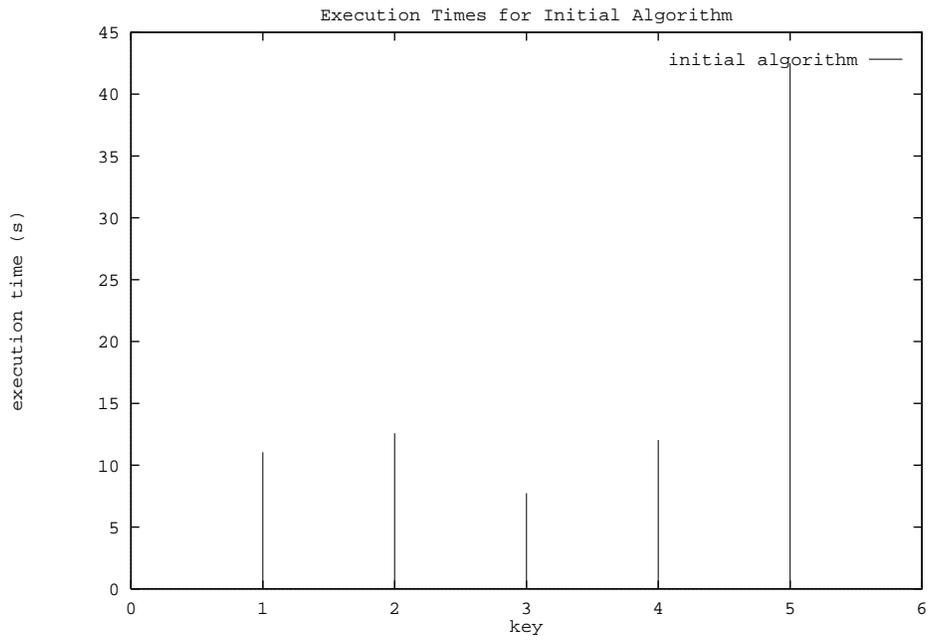

Figure 1: Initial Timing Results

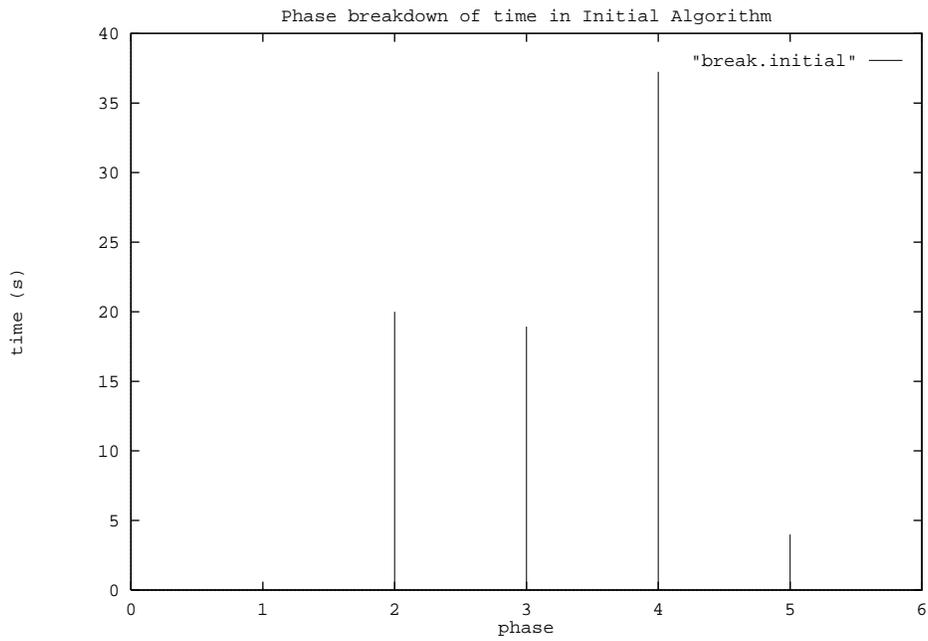

Figure 2: Breakdown for Initial Cumulative Timing Results

### 4.2.1 Selection

The most appropriate course of action was to continue to use the same algorithm for the first four distributions, and to switch to a different algorithm for the fifth distribution. Identifying the fifth distribution (and others related to it) involved examining the output of the bucketing stage. The difficulty with processing the fifth distribution resulted from the clustered nature of the elements. It was impossible to obtain the required resolution for the bucketing to be deemed a success. At the end of the bucketing stage, it was found that the elements of the first 15 cells had significant overlap. That is, a large amount of sorting was still required in order to fully sort the data.

Since the overlap of elements is the identifying feature of such a distribution, an algorithm was developed to gauge the level of overlap and select an alternative sorting algorithm if necessary.

The selection algorithm involved all cells sending a copy of their maximum element to their right neighbour. This was followed by receiving the maximum element from their left neighbour. At this stage, each cell would step through its elements at evenly spaced intervals to determine the number of sample elements which are [8] less than the received maximum. Since the size of the interval was known, this could be used to approximate the number of elements which were overlapping between the two cells. Cells with 65 % or more overlapping elements were deemed to be cells which may have represented a problem to the linear algorithm. If more than two such cells existed, the linear sorting algorithm was abandoned and a more comprehensive algorithm was chosen. The actual selection was achieved by using a call to xy_isum(), such that all cells with 65 % or more overlapping points supplied a 1 to the function, while all remaining cells supplied a 0.

### 4.2.2 Sorting

In all cases where the linear sorting algorithm was determined to be inappropriate, a Batcher's communication pattern was used instead. The performance gain was highly significant. Instead of requiring in excess of 40 seconds to perform the sort, only 20 seconds were needed. It should be noted that the Batcher's sorting algorithm involved replacing only the linear communication pattern used previously. The merge-exchange step remained the same.

There was a minor degradation in the performance of sorting the other distributions, although it was far from significant.

## 4.3 Selection Performance

The performance of the sorting algorithm described in §4.2 was much more acceptable than that of the original algorithm. Figure 3 shows the breakdown of the timing results for the new algorithm. The obvious difference between this plot and that seen in Figure 2 is the significantly smaller peek for phase 4, the cleanup stage. The time taken for the cleanup stage fell below that taken for either the bucket or local sort stages. Another point to note is that the peaks for the other phases of the algorithm remained relatively constant. This supports the notion that the selection process had a relatively small overhead.

---

[8] In this implementation, a sample consisting of 1000 points was used

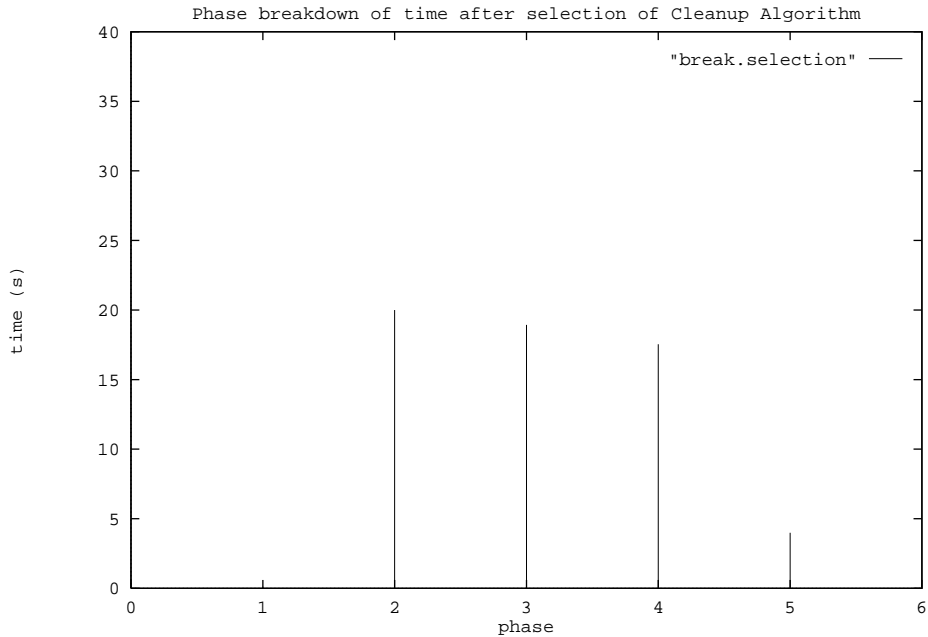

Figure 3: Breakdown for Selection Cumulative Timing Results

## 4.4 Communication

The original implementation of the bucketing algorithm involved a naive global summation of the bucket totals. During the bucketing process, it was necessary to form a global sum on each cell for each of the buckets. The initial implementation used a succession of `xy_isum` calls. For algorithm development, this method proved satisfactory. However, in order to achieve high performance, a more comprehensive global summation algorithm was required. The need for all cells to receive the global sum and the fact that the number of cells in use by the program was a power of two suggested that a global hypercube summation may be appropriate.

The purpose of the hypercube global summation algorithm was to take an array of bucket totals from each cell and sum them elementwise. The communication properties of the algorithm are ideal for use with the AP1000. As seen in Figure 4, the algorithm is extremely "noisy", since every processor is sending at every step. Fortunately, on the AP1000, contention is rarely observed and this case was no exception. Of more importance is the $O(log_2(N))$ property of the algorithm compared to the 8192 `xy_isum()` calls that were made in the original implementation. Being able to complete the summation in six steps was an attractive prospect.

To complement the efficient hypercube communication algorithm, a fast method of performing an elementwise summation on two arrays was required. The algorithm was obvious, however the speed improvement resulting from the use of 8-way loop unrolling was unexpected.

The performance gain from using this method to compute the global sum was significant. Originally, communication on this step was taking in the order of 1.5 seconds. Although this figure includes more than the 8192 calls to `xy_isum`. Use of the hypercube

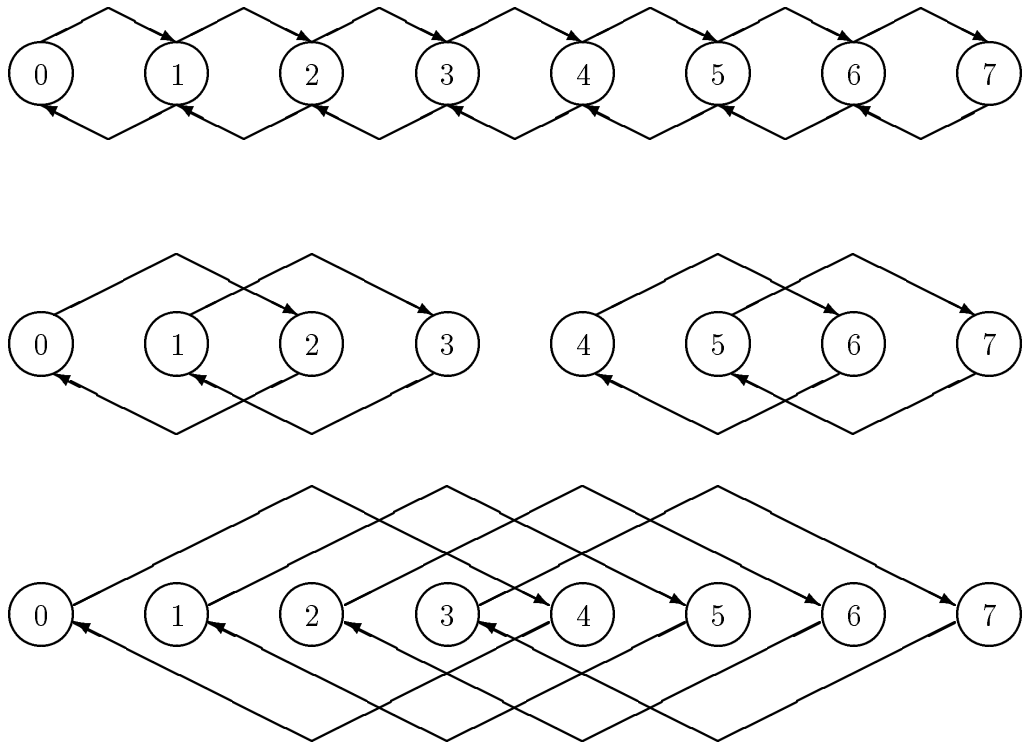

Figure 4: Hypercube Global Summation

algorithm provided an overall gain of 5 seconds across the 5 data distributions.

Since the global summation was required in order for each cell to obtain cumulative bucket totals, an efficient ordering of the algorithm was to perform the cumulative totals locally before the global summation. Thus at the end of the summation, the cumulative bucket totals across all cells were available on each cell.

## 4.5 Radix Sort

As expected, the local sort consumed the most time out of all the phases of the sorting algorithm. As a result, any gains made in this area greatly improved overall performance. The radix sorting algorithm was generally accepted as the fastest algorithm to use under the specified sorting conditions. Improved performance from such an algorithm was achieved through the extensive use of loop unrolling. It was found that unrolling loops by a factor of 8 was beneficial. Further attempts at unrolling appeared not to produce any measurable improvements.

The use of relatively large amounts of static memory appeared to improve performance to some degree. The cost however was less available memory for other parts of the sorting program.

A further performance gain was achieved by noting that most of the elements in the sample data sets were 30 bit numbers. Reducing most comparisons to 30 bits produced a marked performance gain. Obviously the 31 bit case still had to be handled. The cost of this optimisation was a slight loss in performance when handling numbers making use of the full 31 bits.

## 4.6 Merge-Exchange

The optimised radix sort routines inherited from other Computer Science Honours students had significantly greater static memory allocations than was previously used. As a result, the original merge-exchange routine described in §3.4 was unable to execute successfully due to lack of memory. The solution was to replace this routine with an in-place merge-exchange routine. However, in order to use such a routine, all cells were required to have the same number of elements. This resulted in a major restructuring of the sorting algorithm. The revised sorting algorithm is as follows:

1. pre-balance
2. bucketing
3. post-balance
4. radix sort
5. cleanup

### 4.6.1 General Algorithm

As with the original implementation, a negotiation process was required to determine the amount of data to be exchanged between two cells. In this case however, the exchange of data is exact. There is an increased amount of communication as the two cells must exchange sample data elements until an agreement is reached. This was an ideal situation to make use of the `xy_send()` and `xy_recvs()` routines of the AP1000. Once the cells have determined the exact number of data elements to exchange, the exchange takes place. In this situation however, any received data is read directly into the data array, overwriting the data that was sent to the other cell.

With all the necessary data on the local cell in the same array, the problem reduces to a local sort. In this situation however, the data is almost sorted since two sorted lists are stored in the array. As a result, use of standard sorting algorithms described previously is inappropriate. Instead, an algorithm whereby the two sublists stored in the array are treated as separate units is used. Essentially, the two lists are merged in blocks. Memory is required for temporary blocks which store data elements as necessary. The first block from each sublist is copied into temporary locations. These blocks are then merge sorted into another temporary block. When the target block is filled, it is copied to the next available location in the target array. As each source block is consumed, a new block is copied from the source array to take its place. This process continues until the merge-sort is complete.

The memory advantages of this algorithm are significant. Ultimately, the algorithm leads to much larger data sets being able to be sorted.

## 4.7 Bucketing

By using an in-place merge algorithm, the demands on memory were greatly reduced. This provided an opportunity to increase the number of buckets used in the bucketing stage of the algorithm. Increasing the number of buckets from 4096 to 8192 increased the performance of the cleanup stage because the increased resolution allowed a more accurate global sort to be achieved.

## 4.8 Memory Movement

Throughout the entire sorting process, memory is constantly being copied from one location to another. In the case a distributed memory machine, memory is often copied from the system area to user memory. A measurable performance gain was achieved by using more efficient implementations of the `memcpy()` routine. An assembler version of the routine, written by Gordon Irlam, was included into the sorting program. The improvement however, was not as substantial as those obtained by the methods described above.

## 4.9 Final Performance

By using all of the optimisation methods described above, much higher performance was achieved. Figure 5 shows the final timing results for each of the five keys. Key 5 was still the most difficult distribution to sort quickly, however reducing sorting time from in excess of 40 seconds to approximately 13 seconds was a significant gain. Obvious gains were also achieved on the other keys.

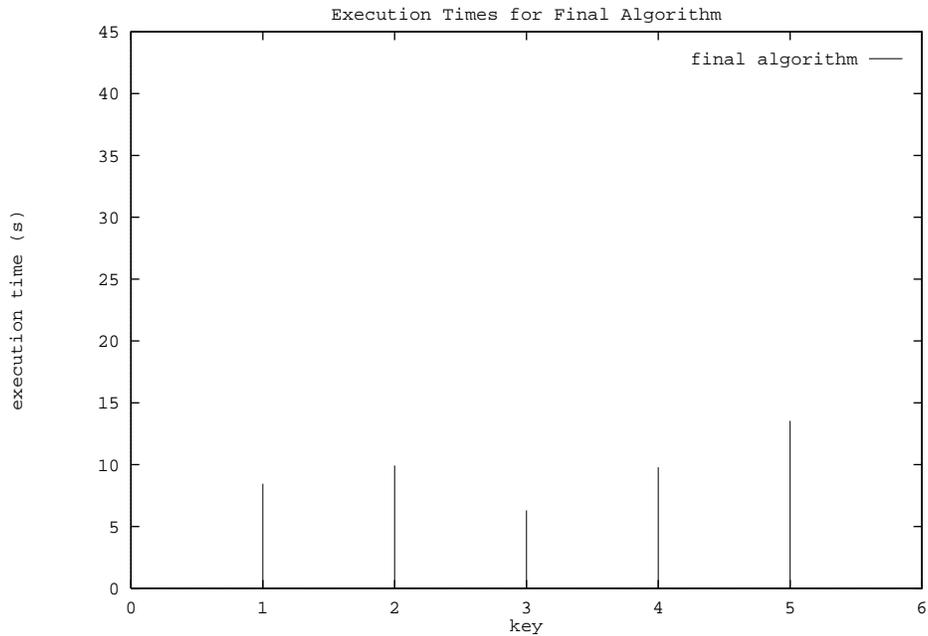

Figure 5: Final Timing Results

The breakdown of the timing results is also of interest when considering the performance of the final algorithm. The most significant attribute of this plot is the fact that the local sort consumes the most time out of all the stages of the algorithm. In the original algorithm, it was the bucketing phase which consumed more time than any other phase of the algorithm. The bucketing phase obviously benefited from the use of the hypercube summation algorithm. The cleanup phase benefited from the use of a Batcher's sort when necessary and the increased number of buckets. However, the time taken for the post-balance phase remained constant. As a result, the post-balance had a greater impact on the overall performance than was originally the case. The performance of the pre-balance remained unchanged.

## 5  Global Radix — an alternative?

As mentioned previously, the bucketing phase of the algorithm described above is very similar to the last pass of a radix sort. That this phase is generally effective and not excessively expensive[9] begs the question of how effective a global radix sort would be. To be worthwhile, a global radix sort must be at least as fast as other methods, hence it must average approximately 8 seconds per run, or complete the five keys in 40 seconds or less[10]

---

[9]Individual timing runs for each key broken down into the separate phases revealed that bucketing generally takes 3.09 seconds.

[10]The best result achieved by the Honours class was a total of 40 seconds for the five keys. The algorithm was evolutionarily convergent with the one described, with the one remaining difference when

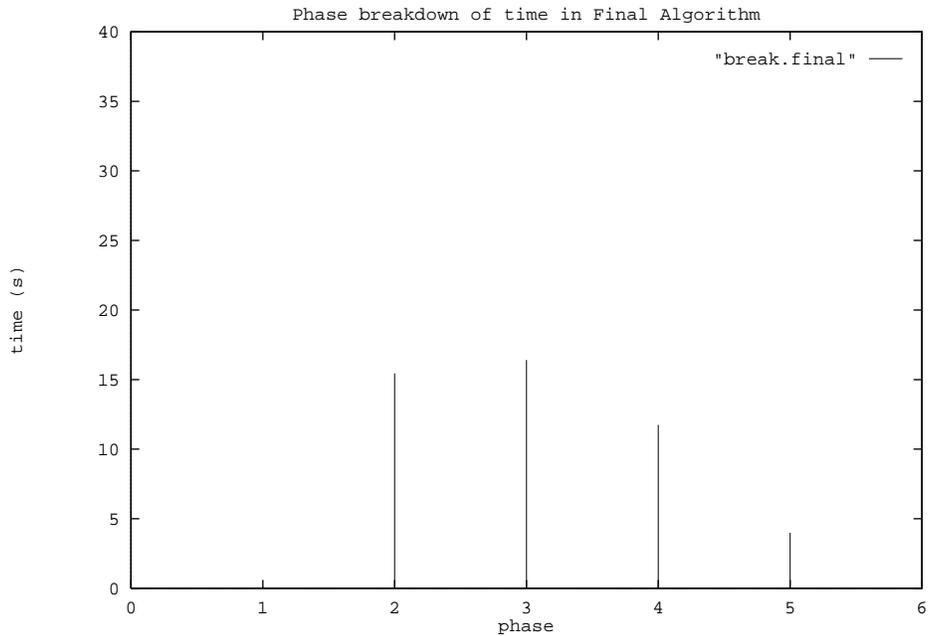

Figure 6: Breakdown for Final Cumulative Timing Results

To do a single pass radix sort on the positive integers would require approximately 8.5 gigabytes of memory. Since each cell of an AP1000 possesses only 16 megabytes, a single pass global radix sort is out of the question. At 3.09 seconds per pass a three pass sort would require in excess of 9 seconds, substantially above the 8 second average. Hence only a two pass sort could possibly be competitive. Each pass of a 2 pass radix sort would however require slightly longer than the bucketing phase's 3.09 seconds. Firstly, more buckets are required, 65 536 on the first pass and 32 768 on the second, compared to only 8 192 being used in the bucketing phase. These extra buckets would require more time to total, both locally and globally. Secondly, the grouping of elements within their buckets must be preserved through the distribution to other cells. At present when elements are received from a cell they are ordered according to the partial sort performed by the sender, and once all elements are received only those from a particular sender are in the partially sorted order, as a whole the received data is effectively unsorted. This problem could be overcome by distributing individual buckets and with the receiver processing them in order (perhaps by message type, although any mechanism would suffice). Unfortunately observation of other implementations using a bucket distribution scheme of similar mechanics to that just mentioned has indicated that performance suffers dramatically. Another possible method to solve this problem would be to send local bucket totals along with the elements so that the receiver could split up and reorder the data as necessary. This method would probably be less expensive than the multi-bucket transmission, but would still entail not only extra messages but at least one extra layer of memory copying in the receive, which is expensive ($>0.6$ seconds) when dealing with up to 4 megabytes of data.

---

implementation ceased, being a perfectly balanced bucketing result.

Assuming identical pre-balance times for a global radix sort and the implemented algorithm, the global radix would need to complete both passes in $8 - 0.0235 = 7.9765$ seconds, or $\approx 3.988$ seconds per pass. Hence the extra buckets must be totaled and the reordering of received data achieved inside of 0.9 seconds per pass. Further investigation, most probably an implementation, would be necessary to determine if this was possible. Unfortunately time has not permitted this, but it would appear that a global radix sort would be competitive with the best reported times for other methods, if not marginally faster.

# 6  Conclusion

The high performance algorithm described in this paper was developed by making use of both the knowledge of the data to be sorted and parallel sorting algorithms. Knowledge of the data to be sorted, 32 bit integers in this case, prompted the use of highly efficient sorting algorithms suitable for this kind of data. In particular, a radix sorting method was used for all serial sorting. Since the serial sort ultimately took the largest amount of time of all the algorithm phases, significant performance benefits were obtained by making use of well know optimisations such as loop unrolling. In addition, other optimisations of a non-parallel nature included the use of more efficient `memcpy()` routines.

The key to performance on a massively parallel machine such as the AP1000 lies in the parallel algorithm. A common practice in parallel algorithm design is to minimise the amount of communication. However, in this situation, the high speed of the AP1000 communication hardware compared to the power of the processing elements allowed for much more scope. By making significant use of communication, algorithms such as that used for the bucket distribution could be made to achieve high performance in terms of execution time. Typically noisy communication patterns such as that used in the Batcher's cleanup phase or the hypercube global summation did not show signs of suffering from contention. As a result, the gains in overall performance far outweighed any communication overhead.